\documentclass[%
 reprint,
superscriptaddress,
 amsmath,amssymb,
 aps,
]{revtex4-2}

\usepackage{graphicx}
\usepackage{dcolumn}
\usepackage{bm}
\usepackage{siunitx}
\usepackage[hidelinks]{hyperref}

\newcommand{\au}{\,\mathrm{a.u.}}
\begin{document}

\preprint{APS/123-QED}

\title{Coincidence measurement of two-photon double ionization of argon through an autoionizing resonance}


\author{Sebastian Hell}
\affiliation{Institute of Optics and Quantum Electronics, Friedrich-Schiller University, 07743 Jena, Germany}

\author{Julian Sp\"athe}
\affiliation{Institute of Optics and Quantum Electronics, Friedrich-Schiller University, 07743 Jena, Germany}

\author{Morten F\o rre}
\affiliation{Department of Physics and Technology, University of Bergen, N-5007 Bergen, Norway} 

\author{Robert Klas}
\affiliation{Institute of Applied Physics, Friedrich-Schiller University, 07745 Jena, Germany}
\affiliation{Fraunhofer Institute for Applied Optics and Precision Engineering, Albert-Einstein-Str. 7, 07745 Jena, Germany}

\author{Jan Rothhardt}
\affiliation{Institute of Applied Physics, Friedrich-Schiller University, 07745 Jena, Germany}
\affiliation{Fraunhofer Institute for Applied Optics and Precision Engineering, Albert-Einstein-Str. 7, 07745 Jena, Germany}
\affiliation{Helmholtz Institute Jena, 07743 Jena, Germany}

\author{Jens Limpert}
\affiliation{Institute of Applied Physics, Friedrich-Schiller University, 07745 Jena, Germany}
\affiliation{Fraunhofer Institute for Applied Optics and Precision Engineering, Albert-Einstein-Str. 7, 07745 Jena, Germany}
\affiliation{Helmholtz Institute Jena, 07743 Jena, Germany}

\author{Robert Moshammer}
\affiliation{Max Planck Institute for Nuclear Physics, 69117 Heidelberg, Germany}

\author{Christian Ott}
\affiliation{Max Planck Institute for Nuclear Physics, 69117 Heidelberg, Germany}

\author{Gerhard G Paulus}
\affiliation{Institute of Optics and Quantum Electronics, Friedrich-Schiller University, 07743 Jena, Germany}
\affiliation{Helmholtz Institute Jena, 07743 Jena, Germany}

\author{Stephan Fritzsche}
\affiliation{Helmholtz Institute Jena, 07743 Jena, Germany}

\author{Matthias Kübel}
 \email{matthias.kuebel@uni-jena.de}
\affiliation{Institute of Optics and Quantum Electronics, Friedrich-Schiller University, 07743 Jena, Germany}
\affiliation{Helmholtz Institute Jena, 07743 Jena, Germany}

\date{\today}

\begin{abstract}
We present coincidence measurements of two-photon double-ionization (TPDI) of argon driven by femtosecond pulses tunable around \SI{26.5}{eV} photon energy, which are obtained from a high-harmonic generation source. The measured photoelectron spectra are interpreted with regard to three TPDI mechanisms. Theoretical predictions are obtained by an approximate model for direct TPDI and atomic structure calculations, which are implemented into a Monte Carlo simulation. The prevailing mechanism involves the excitation and prompt photoionization of an autoionizing resonance in neutral argon. We provide evidence for pronounced electron-electron interaction in this ultrafast ionization process. Furthermore, we show that the dominant TPDI mechanism can be altered by slight tuning of the photon energy. The present work paves the way for scrutinizing and controlling non-linear photoionization in the extreme ultraviolet using table-top sources.
\end{abstract}

\maketitle


 Two-photon double-ionization (TPDI) of atoms is one of the fundamental non-linear processes involving correlated electron dynamics. TPDI may take place if the energy of two photons exceeds the sum of the first and second ionization potentials, placing the process in the extreme ultraviolet (XUV) spectral region. 
 
 TPDI comes in two variants: First, sequential TPDI is favored if the photon energy exceeds the second ionization potential. In this case, the energies of the two photoelectrons are given by the differences between the photon energy and the first and second ionization potentials, respectively. Resonances may give rise to additional sequential pathways, potentially below the second ionization potential.  Second, direct TPDI may take place if the energy of one photon is lower than the second ionization potential,~i.e., sequential TPDI is not possible. Then the excess energy is shared between the photoelectrons \cite{Nikolopoulos2001, Horner2007}. 
 
 When the sequential pathway is possible, it is strongly favored because of the stable intermediate state, which allows essentially unlimited time for the second photon to be absorbed. In contrast, direct TPDI proceeds via a virtual state, requiring both photons to be absorbed within the energy-time uncertainty. Hence, it is only favored if the light intensity is very high or the pulse duration is extremely short \cite{Feist2009}. For the same reason, a strong impact of electron-electron interaction is expected for direct TPDI, which manifests in the energy sharing between both photoelectrons. 
 
 Significant effort has been undertaken to accurately model the electron-electron interaction by predicting the energy sharing ratio \cite{Hu2005,Feist2008,Forre2010,Jiang2015,Chattopadhyay2023}. To this end, also the angular distributions and correlations have been investigated \cite{Fritzsche2008,Pazourek2011}. So far, the vast majority of theoretical work has focused on helium, owing to its simple electronic structure. Nevertheless, predictions exist also for larger atoms \cite{Forre2010,Gryzlova2019, Chattopadhyay2024}.

Accessing the correlated electron dynamics experimentally requires coincidence detection schemes \cite{Dorner2000,Ullrich2003}. While a substantial body of work has been dedicated to its strong-field counterpart \cite{Weber2000,Doerner2002,Bergues2015}, coincidence experiments on TPDI have been hampered due to a lack of suitable lab-based XUV sources. So far, most experiments have been carried out using free-electron lasers \cite{Moshammer2007,Rudenko2008,Kurka2009,Augustin2018,Straub2022}.  These experiments have confirmed the usual predominance of sequential double ionization \cite{Kurka2009}. Nevertheless, also in this case, evidence for electron-electron interaction has been obtained \cite{Augustin2018}. 

Using table-top XUV sources, multiple ionization of atoms has been achieved \cite{Nabekawa2005,Hasegawa2005,Benis2006} and utilized for the measurement of attosecond pulse trains \cite{Tzallas2003,Sekikawa2004} and isolated attosecond pulses \cite{Bergues2018}. However, at the typical low repetition rates of these sources, coincidence experiments have hardly been feasible. Recent advances in laser technology and high-harmonic generation (HHG) sources have made non-linear processes in the XUV accessible using table-top sources with kHz repetition rates \cite{Major2021,Kretschmar2022}. Moreover, it has been shown that the conversion efficiency of HHG can be significantly increased by using short driving pulses in the visible spectral range \cite{Klas2016}. This allows generating XUV pulses with sufficient pulse energy at much higher repetition rates \cite{Comby2019,Klas2021}.

\begin{figure}
    \centering
    \includegraphics[width=\linewidth]{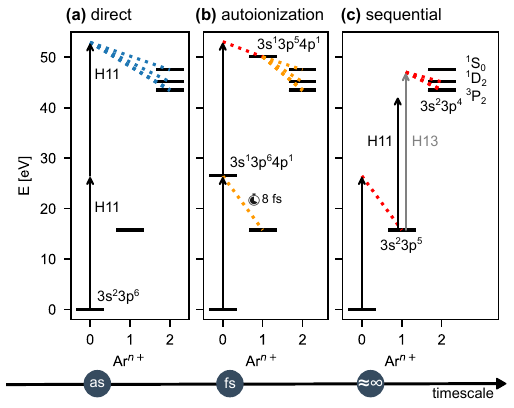}
    \caption{Electronic energy levels of argon and its cations, relevant to the different TPDI mechanisms contributing in our experiment: (a) direct, (b) autoionization, and (c) sequential TPDI. The valence shell electron configurations are indicated next to the energy levels. The vertical arrows represent the absorption of a 11\textsuperscript{th} harmonic photon (H11) with \SI{26.5}{eV} energy (or a H13 photon with \SI{31.3}{eV} energy). The relaxation to electronic states after the absorption of XUV photons is indicated by dotted lines, with lifetimes of autoionizing states denoted by a stopwatch symbol. The line color indicates the type of relaxation process, \textit{e.g.} orange for an Auger decay and blue for the sharing of energy between two photoelectrons. Furthermore, the timescale of the TPDI mechanisms is indicated below each subfigure.}
    \label{fig:level-diagramm}
\end{figure}

Here, we present coincidence measurements of Ar$^{2+}$ and one photoelectron using quasi-monochromatic femtosecond pulses centered at photon energies around \SI{26.5}{eV}, obtained from a \SI{100}{kHz} HHG source \cite{Spaethe2025}. The measured data allows us to test different models of TPDI, ranging from direct to sequential mechanisms, as illustrated in Fig.~\ref{fig:level-diagramm}. The photon energy we chose for our experiment is below the second ionization potential of Ar (\SI{27.6}{eV}), so direct TPDI is the only expected process. 

However, the presence of a window resonance close to the chosen photon energy opens another path for double ionization as depicted in Fig.~\ref{fig:level-diagramm}(b). First, a 3s electron is photoexcited to the 4p state. The resulting [Ne]3s\textsuperscript{1}3p\textsuperscript{6}4p\textsuperscript{1} state of neutral Ar is autoionizing and has a lifetime of $\sim \SI{8}{fs}$ \cite{Sorensen1994}. Absorption of another H11 photon within this lifetime can lead to the removal of a 3p electron. The reached cationic [Ne]3s\textsuperscript{1}3p\textsuperscript{5}4p\textsuperscript{1} state is again autoionizing and decays by emission of an Auger electron into any of the Ar$^{2+}$ ground states. This TPDI mechanism via a window resonance is sequential within the ultrashort lifetime of the intermediate autoionizing state. As will be shown below, the presence of the window resonance may enhance the double ionization probability, even though the single ionization probability is decreased \cite{Sorensen1994}.  In addition, a 6\,\% contribution of the 13th harmonic (H13) at \SI{31.3}{eV} enable sequential TPDI via the ground state of Ar$^+$. 

The experimental setup, including the high-harmonic beamline, will be described in detail elsewhere \cite{Spaethe2025}. Briefly, 30-fs pulses centered at $\SI{515}{nm}$ are obtained from the frequency-doubled output of a post-compressed Yb:glass fiber laser operating at a repetition rate of \SI{100}{kHz}. The central wavelength is tunable in the range from $\approx \SI{512}{520}{nm}$ by slightly tilting the \SI{300}{\mu m} thick beta barium borate crystal used for second harmonic generation. The pulses with their energy adjusted to $\SIrange{60}{80}{\mu J}$ are focused into a gas jet for high-harmonic generation. The generated XUV radiation above \SI{20}{eV} is separated from the residual visible light by passing it through a Al filter with a thickness of \SI{250}{nm} (\SI{150}{nm} for the tuning experiment). The transmitted XUV spectrum is characterized using a parasitic photoelectron time-of-flight spectrometer and contains harmonics of orders 9, 11, and 13. The XUV pulse duration is estimated at $\approx\SI{15}{fs}$ \cite{Klas2021}. The XUV pulses pass several differential pumping stages before entering a reaction microscope \cite{Dorner2000, Ullrich2003}, where a background pressure $<\SI{1e-10}{mbar}$ is achieved. The XUV pulses are back-focused into a cold jet of argon atoms using a suitably coated \cite{Yulin2004} mirror ($f = \SI{75}{mm}$) to effectively monochromatize the reflected XUV light. Ions and electrons are detected in coincidence, with the total count rate kept well below one event per laser pulse, facilitating the measurement of clean ion-electron coincidences. 

\begin{figure}
    \centering
    \includegraphics[width=\linewidth]{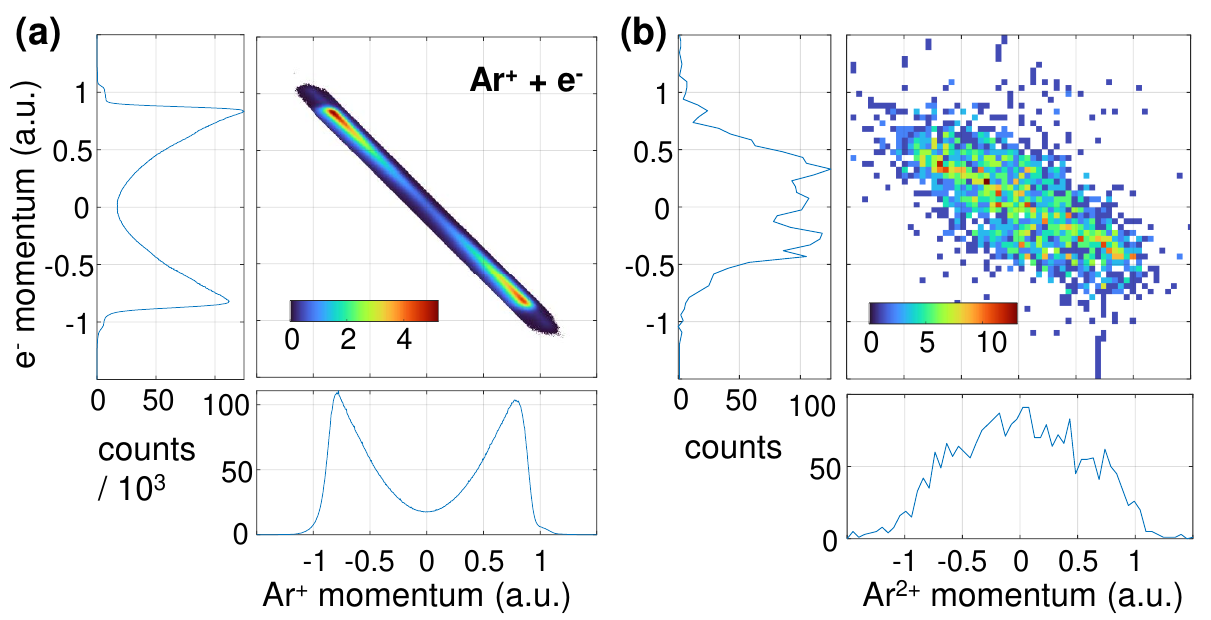}
    \caption{Photoelectron-photoion coincidence spectrum showing the correlations between the momentum components along the XUV polarization for photoelectron and (a) Ar$^+$ ion, or (b) Ar$^{2+}$ ion, respectively.}
    \label{fig:coincidences}
\end{figure}

Figure \ref{fig:coincidences} displays experimental results showing the coincident measurement of singly and doubly charged argon ions together with a photoelectron. For photoelectrons, high resolution ($\Delta p \approx 0.03\au$) is obtained in all three spatial dimensions. Atomic units (a.u.) are used for momentum. For ions, high resolution is achieved in the direction of the XUV polarization only ($\Delta p_{||} \approx 0.07\au$). Along this axis, a condition for momentum conservation is used to unambiguously select true coincidences of Ar$^+$ ions and photoelectrons, meaning that ion and photoelectron originate from the same ionization event. 

In the case of coincident detection of Ar$^{2+}$ and one photoelectron, discrimination based on momentum conservation cannot be used since the second, undetected photoelectron also carries momentum. For this reason, the momenta of Ar$^{2+}$ and one electron are only loosely correlated with a slope of 0.5, as seen in Fig.~\ref{fig:coincidences}(b). Triple coincident events of Ar$^{2+}$ and two photoelectrons are rare in our experiment and thus are disregarded in the analysis. Fortunately, the symmetry of the coincidence plot for Ar$^{2+}$ + e$^-$ indicates negligible contributions from false coincidences. Based on the data recorded for single ionization, the false-coincidence contributions are estimated to be well below \SI{10}{\%}.

\begin{figure}
    \centering
    \includegraphics[width=0.95\linewidth]{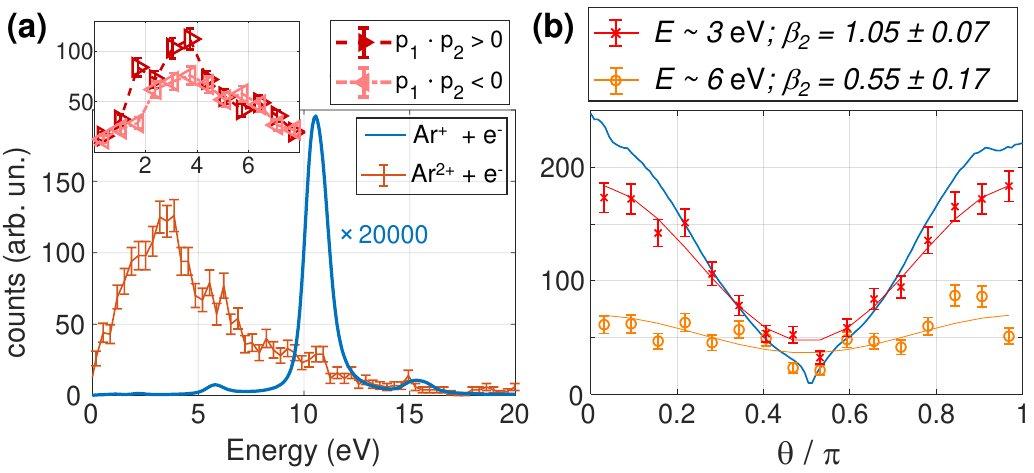}
    \caption{(a) Photoelectron spectra for coincidence events of Ar$^+$ (blue) and Ar$^{2+}$ (red) ions with one photoelectron. The curve for Ar$^+$ was divided by a factor of $20,000$ for visual convenience. The inset shows the spectra of photoelectrons detected in coincidence with Ar$^{2+}$, filtered by whether their momentum components along the XUV polarization are equal ($p_1 \cdot p_2 > 0$) or unequal ($p_1 \cdot p_2 < 0$). (b) Measured photoelectron angular distributions for electrons detected in coincidence with Ar$^+$ (blue) and Ar$^{2+}$ (red / orange). For the latter, two different energy regions are considered, exhibiting angular distributions with different degrees of anisotropy, as displayed in the figure legend.}
    \label{fig:electron_energy_angle}
\end{figure}

The measured photoelectron spectra are presented in Fig.~\ref{fig:electron_energy_angle}. The photoelectron energy distribution recorded for single ionization is dominated by a strong line at \SI{10.7}{eV}, indicating that \SI{90}{\%} of the ionization is due to the absorption of a photon from the 11th harmonic (H11); the contributions of H9 and H13 are \SI{4}{\%} and \SI{6}{\%}, respectively. 

The energy distribution of photoelectrons detected in coincidence with Ar$^{2+}$ differs strongly from the one measured for single ionization. The absence of a pronounced line at \SI{10.7}{eV} shows that sequential TPDI via the ground state of Ar$^+$ [cf.~Fig.~\ref{fig:level-diagramm}(c)] can be ruled out as the main double ionization mechanism in our experiment. We further note that the electron energy spectrum measured for Ar$^{2+}$ is not mirror-symmetric with respect to a certain energy value, as would be expected if a single final state of the ion was reached after double ionization. Indeed, Ar$^{2+}$ possesses a pronounced fine structure splitting, with essentially three different energy levels, as illustrated in Fig.~\ref{fig:level-diagramm}. Due to the contributions from multiple final states the electron energy spectrum is not equivalent to the energy sharing ratio.

The data presented in the inset of Fig.~3(a) indicates a direct imprint of electron correlations. The plot has been generated by first calculating the momentum component of the undetected electron along the XUV polarization, $p_2 = -p_i - p_1$, where $p_i$ is the measured Ar$^{2+}$ ion momentum component along the XUV polarization, and $p_1$ is the corresponding momentum of the detected electron. The energy distribution of the detected electron is plotted separately for events with parallel ($p_1 \cdot p_2 > 0$) and anti-parallel ($p_1 \cdot p_2 < 0$) emission of both electrons. In the range between 2 and 4 eV, electron pairs are more frequently emitted parallel than anti-parallel. This observation contradicts the single-active electron approximation, for which no such correlation is to be expected.

The photoelectron angular distributions for single and double ionization are presented in Fig.~\ref{fig:electron_energy_angle}(b). The angular distributions are quantified by the anisotropy parameter $\beta_2$, which is determined by fitting a second-order Legendre polynomial to the measured angle-dependent yield. 
For single ionization by H11, we find $\beta_2 = 1.38 \pm 0.01$, in agreement with published data \cite{Houlgate1976}. For double ionization, a fit up to fourth order is used. We find significantly different values of $\beta_2$ for different photoelectron energies. In the \SIrange{2}{4}{eV} range, $\beta_2$ is similar to the value observed for Ar$^+$, whereas the photoelectron angular distribution is significantly more isotropic for energies in the \SIrange{5}{8}{eV} range, corresponding to a smaller value of $\beta_2$. The small values for $\beta_4$ ($-0.12 \pm 0.06$ in the \SIrange{2}{4}{eV} range and $-0.22 \pm 0.20$ in the \SIrange{5}{8}{eV} range) are disregarded in the further analysis. The measured $\beta_2$ and $\beta_4$ values are consistent with previous experiments \cite{Kiselev2020}.

\begin{figure}
    \centering
    \includegraphics[width=\linewidth]{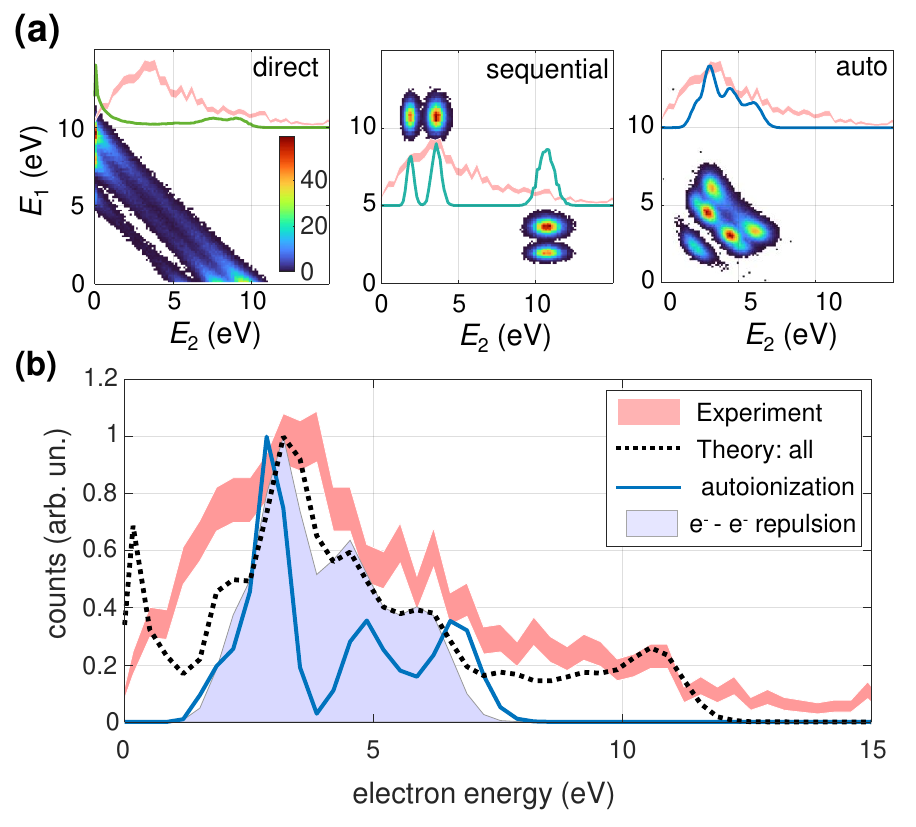}
    \caption{(a) Two-electron energy spectra calculated for the direct, sequential, and autoionization mechanisms, as indicated. The predictions for TPDI through the autoionizing state include the effect of electron-electron repulsion in the continuum. The one-dimensional energy spectra are  drawn in the figure, and compared to the experimental results, for which the 1-sigma confidence interval is plotted (red shaded areas). (b) Photoelectron spectra calculated for the autoionization mechanism, with (blue shaded area) and without (blue line) the effect of coulomb repulsion, and a weighted superposition of all three mechanisms (dotted black line). The computational results are compared to the measured spectrum (red shaded area).}
    \label{fig:models}
\end{figure}

In order to interpret the measured $\mathrm{Ar}^{2+} + \mathrm{e}^-$ data, we implement a Monte Carlo simulation. In the simulation, pairs of photoelectrons with energies $E_1$ and $E_2$ are generated stochastically according to each one of the three different mechanisms illustrated in Fig.~\ref{fig:level-diagramm}: direct, sequential, and autoionization. Each mechanism is implemented such that no free parameters are required. The finite instrument resolution is taken into account by applying a momentum uncertainty of $\delta p = 0.03\au$ in all generated photoelectron pairs. This value is determined by comparing predictions for single-ionization spectra to the experimental results. The Monte Carlo simulations allow us to test the predictions for the three mechanisms by comparing them to our experimental data. In particular, we can evaluate the prediction for the energy sharing ratio, $f_E = \frac{E_1}{E_1+E_2}$, which we present in the form of two-electron energy spectra in Fig.~\ref{fig:models}(a).

For modeling direct TPDI, we use  
\begin{align}
E_1 &= f_E \cdot (2\hbar \omega - I_{\mathrm{P},1} - I_{\mathrm{P},2}^X),\mathrm{~and}   \\      
E_2 &= (1-f_E) \cdot  (2 \hbar \omega - I_{\mathrm{P},1} - I_{\mathrm{P},2}^X).      
\end{align}
Here, $I_{\mathrm{P},1}$, and $I_{\mathrm{P},2}^X$ are the first and second ionization potentials, respectively. 
The number of photoelectron pairs generated for each value of $I_{\mathrm{P},2}^X$ ($X = \text{S, P or D}$) is proportional to the multiplicity of each fine structure level of Ar$_2^+$, \textit{i.e.} 1, 9 and 5, respectively. As discussed in the introduction, the energy sharing parameter $f_E$ is of central interest for characterizing the electron-electron interaction during direct TPDI, and could in principle be retrieved from measured data. Our simulations show that $f_E$ should not be treated as a free parameters as this could lead to the false conclusion that the direct process represents the dominant contribution to the Ar$^{2+}$ yield in our experiment.

The energy sharing parameter $f_E$ is obtained from the model presented in \cite{Forre2010}. Introducing exchange symmetry to the model, 
the singly differential cross section for direct TPDI of argon reads
\begin{align}
\label{TPDI_model_1}
\frac{d\sigma}{dE_1}&=\frac{\hbar^3\omega^2}{4\pi}\left(
\sqrt{g\left(E_1,E_2\right)}+\sqrt{g\left(E_2,E_1\right)}
\right)^2 \\
\label{TPDI_model_2}
g&=
\frac{\sigma_\mathrm{Ar}(E_1+I_\mathrm{P,1} ) \sigma_{\mathrm{Ar}^{+}}(E_2+I_\mathrm{P,2} )}
{(E_1+I_\mathrm{P,1} ) (E_2+I_\mathrm{P,2} ) (E_1+I_\mathrm{P,1}-\hbar\omega)^2},
\end{align}
with $E_1+E_2=2\hbar\omega - I_\mathrm{P,1} - I_\mathrm{P,2}$, and where $\sigma_\mathrm{Ar}$ and 
$ \sigma_{\mathrm{Ar}^{+}}$ are the photoionization
cross sections of Ar and Ar$^+$, respectively. This model has been shown to yield excellent agreement with ab initio calculations for helium. Applied to TPDI of argon at \SI{26.5}{eV}, the model predicts rather asymmetric energy sharing, see Fig.~\ref{fig:models}(a). 

We model sequential TPDI via the Ar$^+$ ground state by assuming that the first ionization step results from the absorption of H11, and the second ionization step results from the absorption of H13. H13 possesses sufficient energy to reach the P and D fine structure states of Ar$^{2+}$, see Fig.~\ref{fig:level-diagramm}(c). The fine structure splitting of $I_{\mathrm{P},1}$ (0.18\,eV) is neglected as it is not resolved in the experimental data for single ionization. The relevant photoabsorption cross-sections have been calculated using the Jena Atomic Calculator (JAC) \cite{Fritzsche2019} and have been used to weight each possible pathway.  The JAC toolbox applies multi-configuration Dirac-Hartree-Fock wave functions \cite{grant2007relativistic} to compute all required cross sections and rates for the coupling of the bound-state electron density to the continuum. These wave functions offer the distinct advantage that they help formulate all ionization and cascade processes right in terms of many-electron amplitudes as suitable for open-shell atoms and ions across the periodic table \cite{fritzsche2024merits}.

The autoionization mechanism of TPDI is modeled analogously to sequential TPDI. All relevant states and transition rates are calculated using JAC. This includes the photoionization from the autoionizing [Ne]3s\textsuperscript{1}3p\textsuperscript{6}4p\textsuperscript{1} state of neutral Ar to the cationic [Ne]3s\textsuperscript{1}3p\textsuperscript{5}4p\textsuperscript{1} manifold, which consists of 18 states, 13 of which can be reached by absorption of H11. The resulting photoelectron energies are mainly in the range from \SIrange{2}{4}{eV}. Furthermore, the energies and decay rates for all 90 Auger lines to the ground states of Ar$^{2+}$ are calculated. The energies are dominantly in the \SIrange{4}{8}{eV} range, and typical lifetimes are few femtoseconds. The corresponding linewidths are taken into account.

The resulting photoelectron spectra for the three mechanisms are presented in Fig.~\ref{fig:models}(b). The predictions for the autoionization mechanism reproduce the position of the peak around \SI{3}{eV} and the shoulder in the \SIrange{4}{7}{eV} range. We emphasize that the latter range is due to Auger electrons, which are characterized by a near-isotropic angular distribution; the peak around \SI{3}{eV} is due to photoelectrons, which tend to have a more anisotropic angular distribution. Hence, the observation of different photoelectron angular distributions in the aforementioned energy ranges, see Fig.~\ref{fig:electron_energy_angle}(b), lend support to the autoionization mechanism. However, the well-defined lines in the predicted photoelectron spectrum are not observed experimentally. 


In the following, we explore whether post-collision interactions in the form of electron-electron repulsion may wash out the measured photoelectron spectra for double ionization. It is expected that electron pairs emitted via the autoionization mechanism, which involve ultrafast Auger-Meitner decay, are more strongly affected than electron pairs emitted via the sequential mechanism, which involves the stable ground state of Ar$^+$. Notably, in the autoionization mechanism, the Auger electron with an energy in the $\SIrange{4}{7}{eV}$ range leaves the atom within only a few femtoseconds after the $\SIrange{2}{4}{eV}$ photoelectron. This allows for significant energy exchange between the two electrons. Further, the tendency for electron pairs to be emitted in the same direction, see inset of Fig.~\ref{fig:electron_energy_angle}(a), may enhance the effect of electron-electron repulsion. The effect of the Coulomb interaction between the Auger electron and the photoelectron is modeled in a second Monte Carlo simulation. In this simulation, pairs of electrons with randomly chosen initial positions, on a sphere of the size of the valence shell of Ar \cite{clementi1967atomic}, and random emission directions according to the measured angular distributions are created. The two electrons are released with a random time delay according to an exponential decay with the calculated and appropriately weighted expectation value of the Auger lifetime ($\approx \SI{6}{fs}$). The classical trajectories of these electron pairs are calculated and the energy exchanged by Coulomb repulsion, while requiring conservation of the total energy, is evaluated. The resulting probability distribution for the energy exchanged between the electrons is used to adjust the momenta of each electron pair in the primary Monte Carlo simulation. The energy exchanged between the electrons pairs leads to the off-diagonal stripes visible in the two-electron spectra shown in Fig.~\ref{fig:models}(a). The projection on one energy axis yields the blue-shaded area plotted in Fig.~\ref{fig:models}(b), which agrees rather well with the experimental data in the range from \SIrange{2}{7}{eV}. %


However, significant discrepancies exist between experimental data and the predictions for the autoionization mechanism at energies below $~\SI{2}{eV}$ and above $\SI{7}{eV}$. Remarkably, these are the energy regions most dominantly populated in the electron spectra predicted for the direct TPDI model with asymmetric energy sharing. 

In order to test whether sequential and direct mechanisms can explain the gaps observed between the measured and predicted electron spectra from the autoionization model, we perform an optimization search using a computational routine. The algorithm calculates the variance between the experimental data and a predicted spectrum which consists of weighted contributions from all three mechanisms. The weights are varied with the goal of minimizing the variance. This procedure results in the dotted black line plotted in Fig.~\ref{fig:models}(b). It is obtained by weighting the spectra for autoionization, sequential (involving H13), and direct mechanisms by \SI{51}{\%}, \SI{16}{\%}, and \SI{33}{\%}, respectively.

\begin{figure}
    \centering
    \includegraphics[width=\linewidth]{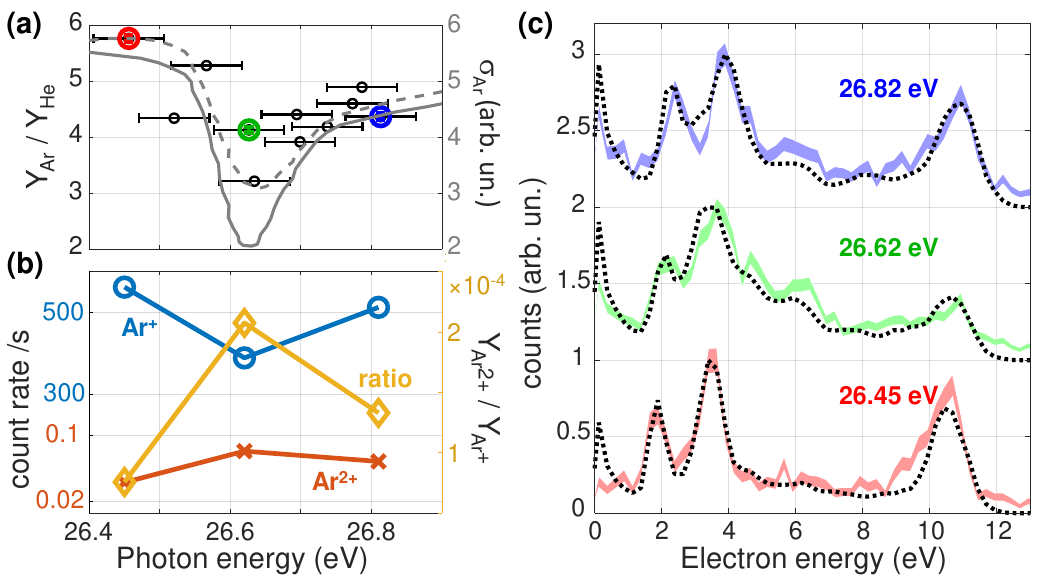}
    \caption{(a) Photon energy calibration based on the ratio of the ion count rate obtained for Ar vs that obtained for He. The error bars represent an uncertainty of \SI{1}{nm} in the determination of the central wavelength of the driving laser. The photon energies used for TPDI measurements are highlighted by colored circles. The photoionization cross-section reported by Sorensen \cite{Sorensen1994} is plotted along with the data points (solid gray line). The dashed gray line is obtained by convolution of the photoionization cross-section with the estimated spectral bandwidths of the 11th harmonic. (b) Coincidence count rates for single and double ionization of Ar (left axis), and their ratio (right axis), obtained for three different photon energies. (c) Photoelectron spectra measured in coincidence with Ar$^{2+}$, for three different photon energies, as indicated (colored shaded areas). The curves are offset vertically for visual convenience, and each one is normalized to its maximum. The dotted lines represent the best fit based on a superposition of the three TPDI mechanisms.}
    \label{fig:tuning}
\end{figure}

In an additional experiment, we control the relative contributions of the different TPDI mechanisms by tuning the photon energy across the resonance. This is achieved by tuning the central wavelength of the driving laser pulses in the range $\SI{512}{nm} \leq \lambda_0 \leq \SI{520}{nm}$, where  sufficient pulse energy for HHG is obtained. Assuming that the photon energy of H11 is given by $E_{11} = 11 hc/\lambda_0$, this corresponds to an expected tuning range spanning from \SIrange{26.2}{26.7}{eV}. This range is sufficient to avoid the window resonance by red-tuning of the photon energy.

The tuning range of H11 is verified in a calibration experiment. The photoionization yield of both argon and helium at fixed gas pressures is measured for three minutes each. Since the photoionization cross-section of helium is essentially flat around \SI{26}{eV}, the ratio of the detected Ar$^+$ / He$^+$ signal is approximately proportional to the photoionization cross-section of argon. Figure \ref{fig:tuning}(a) shows the results of the calibration measurement. The ratio of the ionization yield exhibits a minimum near the center, revealing the precise position of the window resonance. The results indicate that our tuning range is sufficient to scan the photon energy of H11 across the resonance. Thus, the energy of H11 is, in fact, blue-shifted by approximately \SI{0.2}{eV}, relative to the expected tuning range. Such blue-shift is known to arise from the effect of free electrons, see e.g.~\cite{Altucci1999}. The experimental data points plotted in Fig.~\ref{fig:tuning}(a) have been shifted with respect to the expected values. 

Our measurements of the ion yield ratios reveal a much shallower minimum than the one reported for the photoionization cross-section reported in Ref.~\cite{Sorensen1994}. This discrepancy is explained by the spectral bandwidth of the H11 pulse: the cross-section is convoluted with a Gaussian whose width corresponds to the spectral bandwidth of a 15\,fs pulse at \SI{26.5}{eV}. The curve resulting from this convolution (dotted line) agrees well with the observed modulation depth of the yield ratio. However, the data points scatter significantly around the expected trend, suggesting that the actual photon of the harmonics may vary as the central wavelength is changed.

After calibration of the photon energy, we pick three driving wavelengths (519, 515, and \SI{512}{nm}) corresponding to H11 energies below, at and above resonance and carry out TPDI measurements for approximately 20 hours each. The data points for these wavelenghts are highlighted in Fig.~\ref{fig:tuning}(a). Note that further adjustment of the position within the errorbars could be made on the basis of the measured Ar/He ratio; this has not been done for the sake of clarity. Figure \ref{fig:tuning}(b) shows that the single and double ionization rates vary significantly when tuning the photon energy across the resonance. While the Ar$^+$ rate exhibits a minimum on resonance, the Ar$^{2+}$ rate has a maximum.

The photoelectron spectra, recorded in coincidence with Ar$^{2+}$ ions, are presented in Fig.~\ref{fig:tuning}(c). In the off-resonant spectra (blue and red), pronounced peaks are observed at $\approx$ \SI{2}{eV}, $\approx$ \SI{4}{eV}, and \SI{10.7}{eV}. These lines are assigned to sequential TPDI, cf.~Fig.~\ref{fig:models}(a). The ionization of neutral Ar by H11 produces the \SI{10.7}{eV} photoelectron, while the subsequent ionization of the Ar$^+$ ground state by H13 yields the lines at $\approx$ \SI{2}{eV}, $\approx$ \SI{4}{eV}, corresponding to the two lower-lying fine-structure levels of Ar$^{2+}$. In the data measured on-resonance, the \SI{10.7}{eV} line is significantly weaker, and the signal is dominated by a broad distribution ranging from $\approx$\SIrange{2}{7}{eV}. This agrees with the observations of the initial experiment (see Fig.~\ref{fig:electron_energy_angle} (a)) and suggests that autoionization represents the main pathway for TPDI in the on-resonant case. 

The experimental data are fitted with the optimization routine to find the relative contributions of the three mechanisms. In all three cases, there is good agreement between experimental results and theoretical fit. The results of the optimization procedure are summarized in table \ref{tab:results}. 
The fit results are consistent with the expectation that sequential TPDI is the predominant mechanism off-resonant, while autoionization is the strongest contribution on-resonance. The larger contribution of sequential TPDI compared with the initial experiment is attributed to a larger contribution of H13 in the tuning experiment (9\% for the on-resonance case).


\begin{table}
    \centering
    \begin{tabular}{l|l|r|r|r}
             &  & \SI{26.45}{eV} & \SI{26.62}{eV} & \SI{26.82}{eV}\\
             \hline
      Fit result (\%)   & sequential & 48 & 23 & 37\\
         & direct & 28 & 34 & 36\\
         & auto & 24 & 43 & 27 \\
         \hline        
      Effective   & sequential & 5 & 13 & 9\\
    cross-section  & direct & 8 & 8 & 8\\
       (GM)  & auto & $20\pm 9$ & $80 \pm 11$ & $38 \pm 10$\\
         & total  & $33\pm 9$ & $101 \pm 11$ & $55 \pm 10$ \\ 
         \hline                 
        Measured ratio & Ar$^{2+}$/Ar$^+$ & $0.8\cdot 10^{-4}$ & $2.1\cdot 10^{-4}$ & $1.3\cdot 10^{-4}$\\
    \end{tabular}
    \caption{The fit results for the contributions of the three mechanisms to the total double ionization yields are compared with the effective total double ionization cross-sections, given in units of G\"oppert-Meyer.}
    \label{tab:results}
\end{table}

For quantitative analysis of the fitting results, we calculate the absolute two-photon absorption cross-sections for all three mechanisms. 
Table \ref{tab:results} summarizes the \emph{effective} cross-sections, which are obtained as follows: based on the cross-sections of the individual ionization steps obtained with JAC, and the estimated pulse duration, the absolute cross-section for the sequential mechanism is $\sigma_\mathrm{SDI}^{(2)} = \SI{150}{GM}$ ($\SI{1}{GM} = \SI{e-50}{cm^4 s}$), in the absence of the window resonance. This value is weighted with the share of the 13th harmonic in the yield of Ar$^+$ + photoelectron coincidences. The absolute cross-section for the direct pathway is calculated as $\sigma_\mathrm{direct}^{(2)} = \SI{8}{GM}$, using the model described in Ref.~\cite{Forre2010}. Finally, summing over all possible pathways for the autoionization mechanism results in a cross-section of $\sigma_\mathrm{auto}^{(2)} =  \SI{140}{GM}$. This value needs to be weighted by the ratio of spectral overlap between the window resonance and the incident XUV spectrum. In our tuning experiment, this ratio is controlled. Using the determined positions of the HHG spectra and their uncertainty, we calculate spectral overlap ratios of \SI{15 \pm 5}{\percent}, \SI{57 \pm 8}{\percent}, and \SI{28\pm 7}{\percent}, respectively.

The values obtained for the effective cross-section confirm the observed trend that the double ionization signal increases significantly on-resonance, due to the contributions of the autoionization mechanism. However, comparing the fit results with the calculated effective cross-sections indicates that the optimization routine underestimates the contribution of autoionization in favor of direct TPDI.
 One possible cause of this discrepancy is that our simple model of energy exchange between electrons does not accurately capture electron correlation dynamics. However, it is also possible that the effective cross-section for direct TPDI near-resonance is underestimated because the modelling of direct TPDI does not take bound resonances into account.  In either case, our results suggest that advanced multi-electron techniques are required to accurately capture ultrafast correlated electron dynamics.  While there is limited agreement with the fitting results, the calculated total cross-sections reasonably match the boosted double ionisation probability measured at resonance.


In conclusion, we have presented coincidence measurements of TPDI of argon by tunable, quasi-monochromatic (i.e. $\approx 90\%$ of all photons) around \SI{26.5}{eV} radiation obtained from a high-harmonic source. The data is suitable to rigorously test models for different mechanisms of TPDI. At resonance, the best agreement is found for a semi-sequential mechanism involving the excitation of an autoionizing resonance. We provide evidence for correlation in the emission direction of the two electrons. Moreover, the comparison of experimental and computational results suggest that the two emitted electrons interact with each other due to the ultrafast nature of the double ionization mechanism. The accurate treatment of such electron-electron correlations represents a challenge for theories beyond the (mean-field) single-active electron approximation. In future experimental work, near-resonant excitation and ionization dynamics might be manipulated by a combination of XUV tuning and dynamic Stark control exerted by an optical laser. 
Taking a step back, the possibility of conducting coincidence experiments on non-linear photoionization using table-top XUV sources opens up numerous new opportunities for ultrafast science, including XUV-pump, laser-probe and all-XUV pump-probe spectroscopy of molecular dynamics.  \\

\begin{acknowledgments}
We thank Th. Weber, F. Ronneberger and Roentdek for technical support. Funding by the DFG under project no. 437321733 is acknowledged. R.K., J.R. and J.L. acknowledge funding by the BMBF under project EXSAM (13N16673).
\end{acknowledgments}

\bibliography{tpdi}

\end{document}